\documentclass[12pt]{article}

\usepackage{epsfig}
\usepackage{graphicx}
\usepackage{array}
\usepackage{multirow}
\usepackage{a4wide}

\title{Coupling Lattice Boltzmann and Molecular Dynamics models 
       for dense fluids}

\author{A.~Dupuis, E.M.~Kotsalis and P.~Koumoutsakos \\        
        \footnotesize{Computational Laboratory, ETH Zurich, CH-8092, Switzerland.}}

\begin{document}

\maketitle

\begin{abstract}
We propose a hybrid model, coupling Lattice Boltzmann  and  Molecular
Dynamics models, for the simulation of dense fluids. Time and length scales are
decoupled by using an iterative Schwarz domain decomposition
algorithm. The MD and LB formulations communicate via the exchange of
velocities and velocity gradients at the interface. We validate the
present LB-MD model in simulations of flows of liquid argon past and
through a carbon nanotube. Comparisons with existing hybrid algorithms
and with reference MD solutions demonstrate the validity of the
present approach.
\end{abstract}

\newcommand{\pos}{{\mathbf{r}}}
\newcommand{\dt}{{\Delta t}}
\newcommand{\dr}{{\Delta \pos}}
\newcommand{\vi}{{\mathbf{v}_i}}
\newcommand{\vm}{{v}}
\newcommand{\g}{{\mathbf{g}}}
\renewcommand{\u}{{\mathbf{u}}}
\newcommand{\eq}[1]{equation (\ref{#1})}
\newcommand{\Eq}[1]{Equation (\ref{#1})}
\newcommand{\fig}[1]{fig.~\ref{#1}}
\newcommand{\Fig}[1]{Fig.~\ref{#1}}
\newcommand{\teq}{\theta_{eq}}

\section{Introduction}

The advent of nanotechnology provides us today with nanoscale devices
that can affect flow phenomena that are  important for several 
technological applications.  Examples include microfluidic channels with nanopatterned walls ,
biosensors embedded in aqueous environments~\cite{Chen:2003,Li:2003c,Lin:2004b,Zheng:2003},
or bluff bodies with superhydorphobic surfaces~\cite{watanabe:03}.
 
The detailed investigation of flow physics at the nanoscale 
has been pioneered by Koplik \cite{koplik:1995} 
using  Molecular Dynamics (MD) models. 
When nanoscale devices are active parts
of micro and macroscale systems, a hybrid approach is required to
integrate atomistic simulations with computational methods suitable
for larger scales. A number of hybrid models coupling atomistic to
continuum descriptions of dense fluids have been
proposed~\cite{oconnell:95, hadjiconstantinou:97, flekkoy:00, nie:04,werder:05,koumoutsakos:05}.

O'Connell and Thompson~\cite{oconnell:95} coupled an atomistic with a
continuum system where the average momentum of the overlap particles
is adjusted through a boundary force. Flekk{\o}y \textit{et
  al.}~\cite{flekkoy:00} presented a model based on direct flux
exchange between atomistic and continuum regions.  Hadjiconstantinou
and Patera~\cite{hadjiconstantinou:97} proposed to decouple time
scales by using the Schwarz domain decomposition method coupling an
atomistic to a continuum description of a fluid. Convergence to a
steady state solution is achieved through alternating iterations
between steady state solutions within the atomistic and continuum
subdomains. Nie \textit{et al.}~\cite{nie:04,nie:04b} employed a
domain decomposition algorithm to simulate Couette flow over a
nanopatterned surface, as well as a cavity flow where the singularity
at the corners between static and moving walls was described
atomistically. In these simulations a singular boundary force is
employed in order to compensate for the elimination of periodicity in
the MD system. Werder \textit{et al.}~\cite{werder:05} proposed an
algorithm using the alternating Schwarz method to couple an MD model
to an incompressible Navier-Stokes (NS) finite volume solver. They
proposed a novel boundary force, based on the physical characteristics
of the fluid that is being simulated. They reported on simulations of
flows past a CNT and noticed average departures from reference MD
simulations of the order of $4\%$. 

Here we extend the model proposed in  \cite{werder:05}  by replacing the
finite volume solver that was provided by a commercial software package (STAR-CD) 
by a Lattice
Boltzmann (LB) method~\cite{succi-book:01} solving the incompressible
NS equations. The proposed extension aims to take advantage of the
mesoscopic modeling inherent in LB simulations and to allow for a
broader geometric flexibility than the one allowed by the Finite
Volume solver. In addition, in  the present 
work we enhance the exchange between atomistic and
continuum domains by not only communicating velocities as
in~\cite{werder:05} but also velocity gradients.

The paper is organized as follows. In section~\ref{sec:model}, we
describe the hybrid model by outlining the MD and LB methods and describe
their coupling. Results of liquid argon flows past and through a CNT
are presented in section~\ref{sec:results}. We compare the flows to
the reference MD solutions and discuss the computational efficiency of
the hybrid model and  conclude  in section~\ref{sec:conclusion}.

\section{The LB-MD model}
\label{sec:model}
In the present hybrid method an MD model describes the flow in the vicinity of a
carbon nanotube (CNT) while a LB approach is used to simulate the behaviou of the 
continuum system away from the CNT.
\subsection{Molecular dynamics}

The atomistic region is described by  MD
simulations where the positions $\mathbf{r}_i=(x_i,y_i,z_i)$ and
velocities $\mathbf{u}_i=(u_{x,i},u_{y,i},u_{z,i})$ of the particles
evolve according to Newton's equations of motion
\begin{equation}
\frac{d}{dt}\mathbf{r}_i = \mathbf{u}_i 
\quad 
\textrm{and} 
\quad
m_i\frac{d}{dt}\u_i=\textbf{F}_i=- \sum_{j\neq i} \nabla U(r_{ij})
\end{equation}
where $\textbf{F}_i$ and $m_i$ are the force and mass of particle $i$,
and $r_{ij}$ is the distance between the particle $\mathbf{r}_i$ and
$\mathbf{r}_j$. Here we consider Lennard-Jones (LJ) model of argon interacting
with CNTs. The potential $U(r_{ij})$ is defined as
\begin{equation}
U(r_{ij}) = 4 \epsilon_{AB} \left[ \left( \frac{\sigma_{AB}}{r_{ij}} \right)^{12} - 
\left( \frac{\sigma_{AB}}{r_{ij}} \right)^{6} \right] + U_b(r_w,\rho,T)
\label{eq:newton}
\end{equation}
where $\epsilon_{AB}$ and $\sigma_{AB}$ are energy and length
parameters, $A$ and $B$ denote species. The LJ interaction parameters
for argon-argon and argon-carbon interactions are respectively
$\epsilon_{ArAr}=0.9960$ kJmol$^{-1}$, $\sigma_{ArAr}=0.3405$ nm,
$\epsilon_{ArC}=0.5697$ kJmol$^{-1}$, $\sigma_{ArC}=0.3395$ nm. The
term $U_b$ is the boundary potential and accounts for the interaction
of the boundary region with the surrounding medium. It is further
described in~\cite{werder:05}. The CNT is modeled as a rigid structure
to facilitate the investigation of the flow of argon. All interaction
potentials are truncated for distances beyond a cutoff radius
$r_c=1.0$ nm. The equations of motion (\ref{eq:newton}) are integrated
using a leap-frog scheme with a time step $\delta t=10$ fs.

A desired velocity $\mathbf{u}_d$ is enforced by relaxing the center
of mass velocity $\mathbf{\bar{u}}_k=1/N_k \sum_{i\in k} \mathbf{u}_i$
of the $N_k$ particles within a cell $k$ towards $\mathbf{u}_d$
according to a parameter $\lambda$. The velocities $\mathbf{u}_i$ of
the particles in the cell $k$ are updated as
\begin{equation}
\mathbf{u}_i=\mathbf{u}_i+\lambda (\mathbf{u}_d - \mathbf{\bar{u}}_k).
\end{equation}

The slower the velocity relaxes towards its desired value the smaller
the amount of perturbations introduced in the system. On the other
hand choosing a small value of $\lambda$ increases the number of
iterations to reach equilibrium and thus decreases the computational
efficiency. Here we choose as a relaxation parameter $\lambda=0.1$.

\subsection{Lattice Boltzmann model}

The continuum hydrodynamics are described by the incompressible
Navier-Stokes equations
\begin{eqnarray}
\frac{\partial \u}{\partial t} + (\u \cdot \nabla) \u & = & 
  - \nabla p/\rho
  + \nu \nabla^2 \u + \mathbf{g}, 
\label{eq:ns1} \\
\nabla \cdot \u & = & 0
\label{eq:ns2} 
\end{eqnarray}
where $\u$ is the fluid velocity, $p$ the pressure, $\rho$ the density
and $\mathbf{g}$ a body force. We use $\mathbf{g}$ to enforce
Dirichlet boundary conditions, see discussion below.

We solve the equations of motion (\ref{eq:ns1}) and (\ref{eq:ns2}) by
using a Lattice Boltzmann algorithm~\cite{succi-book:01}. This
approach follows the evolution of particle distribution functions
$f_i$ on a $d$-dimensional regular lattice with $z$ links at each
lattice point $\pos$. The label $i$ denotes velocity directions and
runs between $0$ and $z$. $DdQz+1$ is a standard lattice topology
classification. The $D3Q15$ lattice we use here has the following
velocity vectors $\vi$: $(0,0,0)$, $(\pm 1,\pm 1,\pm 1)$, $(\pm
1,0,0)$, $(0,\pm 1, 0)$, $(0,0, \pm 1)$ in lattice units.

\noindent The Lattice Boltzmann dynamics are given by
\begin{equation}
f_i(\pos+ \dt \vi,t+\dt)=f_i(\pos,t)+\frac{1}{\tau}\left(f_i^{eq}(\pos,t)-f_i(\pos,t)\right) + \dt g_i
\label{eq:lbDynamics}
\end{equation}
where $\dt$ is the time step of the simulation, $\tau$ the relaxation
time. The equilibrium distribution function $f_i^{eq}$ is a function
of the density $\rho$ and the fluid velocity $\u$ defined as
\begin{equation}
\rho=\sum_{i=0}^z f_i \quad , \quad \rho\u=\sum_{i=0}^z f_i\vi +\frac{\dt}{2} \mathbf{g}.
\label{lb:velocity}
\end{equation}

\noindent The equilibrium distribution function is chosen as
\begin{equation}
f_i^{eq}(\pos,t) = w_i \rho \left( 1 + \frac{\mathbf{v}_i\cdot\mathbf{u}}{c_s^2} + 
\frac{(\mathbf{v}_i\cdot\mathbf{u})^2}{2c_s^4} - \frac{\mathbf{u}^2}{2c_s^2} \right)
\label{eq:feq}
\end{equation}
where $c_s=1/\sqrt{3}$ is the speed of sound and $w_i$ are weights
chosen as $w_0=4/9$, $w_i=1/9$ for $i=1-6$ and $w_i=1/72$ for
$i=7-14$. The forcing term is defined as~\cite{guo:02}
\begin{equation}
g_i=\left( 1-\frac{1}{2\tau} \right) w_i \rho \left( \frac{\textbf{v}_i-\textbf{u}_i}{c_s^2} +
    \frac{(\mathbf{v}_i \cdot \mathbf{u}_i)}{c_s^4}\mathbf{v}_i \right) \cdot \mathbf{g}.
\end{equation}

\noindent
Performing a Chapman-Enskog  expansion on the LB
dynamics~\cite{BC-livre} shows that equations (\ref{eq:ns1}) and
(\ref{eq:ns2}) are recovered with a kinematic viscosity expressed as
\begin{equation}
\nu=\frac{(\dr)^2}{\dt} \frac{1}{3} (\tau-\frac{1}{2})
\label{eq:visco}
\end{equation}
where $\dr$ is the lattice spacing.

We enforce Dirichlet boundary conditions using a local forcing term
$\mathbf{g}$. The governing equation  \Eq{eq:ns1} is rewritten as
\begin{equation}
\frac{\partial \u}{\partial t} = \frac{\u^*(\pos,t+\dt) - \u(\pos,t)}{\dt} 
                = - (\u \cdot \nabla) \u - \nabla p 
                + \nu \nabla^2 \u = RHS
\label{eq1}
\end{equation}
where $\u^*$ is the velocity at time $t+\dt$ with no forcing term
considered. Including a forcing term leads to
\begin{equation}
\frac{\u^d(\pos,t+\dt) - \u(\pos,t)}{\dt} = RHS + \g(\pos,t)
\label{eq3}
\end{equation}
where $\mathbf{u}^d$ is the desired velocity. Subtracting equations
(\ref{eq1}) from (\ref{eq3}) leads to an expression for the forcing
term
\begin{equation}
\g(\pos,t) = \frac{\u^d(\pos,t+\dt) - \u^*(\pos,t+\dt)}{\dt}.
\label{eq:forcingTerm}
\end{equation}

It is worth noting that evaluating $\u^*$ within an LB method consists
only of performing a streaming step. Indeed by construction of the
equilibrium distribution function ($\rho \u = \sum f_i^{eq}\vi$,
see~\cite{BC-livre} for details), the second term of the right hand
side of \eq{eq:lbDynamics} does not change the velocity $\u$.

\subsection{Hybrid model}

\begin{figure}
\begin{center}
\epsfig{file=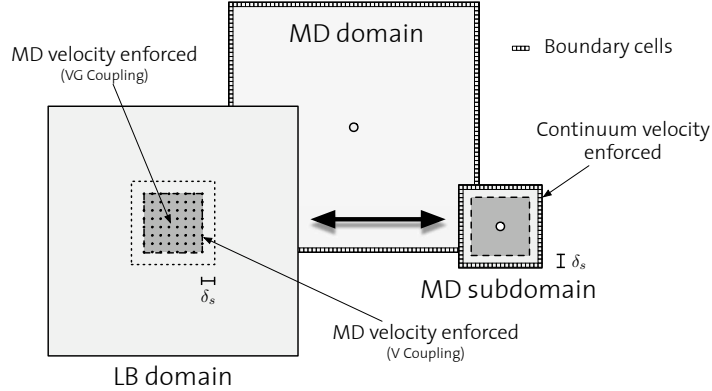,width=10cm}
\end{center}
\caption{Domain decomposition. The dark gray area is computed within
  the MD subdomain and (VGC) enforced within the LB domain or (VC)
  only enforced along a strip depicted by the dashed square. Converged
  solution is obtained by alternating iterations in the LB and MD
  domains.}
\label{fig:decomposition}
\end{figure}

We use a domain decomposition algorithm to couple an MD description of
a dense fluid with an LB model solving the incompressible NS
equations. The computational domain is decomposed into two overlapping
regions of an LB domain and an MD subdomain. \Fig{fig:decomposition}
shows the Schwarz decomposition used to converge to a steady state
solution by alternating iterations between steady state solutions in
the LB domain and MD subdomain.

A Schwarz iteration  $t_c$ consists of computing the continuum velocity
field $\u_c(t_c)$ with boundary conditions set by the previous
atomistic cycle $\u_a(t_c-1)$ and an external boundary condition that
depends on the system considered. Then, $\u_c(t_c)$ is used to set the
boundary condition for computing $\u_a(t_c)$.

MD velocities are sampled in cells of same size as in the LB domain
and are enforced on the continuum according to two coupling methods
using \eq{eq:forcingTerm}. The first approach corresponds to the one
used by Werder \textit{et al.} and is to impose MD velocities within a
one cell wide strip located at a distance $\delta_s$ of the MD
subdomain, see \fig{fig:decomposition}. This velocity coupling (VC)
approach does not enforce velocity gradients implying that the
geometry of the system and the external boundary conditions dictate
whether the hybrid solution evolve into a good approximation of the
reference solution. To alleviate this issue we propose an enriched
coupling method which enforces velocities, and implicitly velocity
gradients (VGC) by imposing MD velocities on every common cell except
within a strip of width $\delta_s$ close to the boundary (see
\fig{fig:decomposition}). We shall observe below that using the VGC
approach leads to closer match with MD reference solutions than when
using the VC method.

We use the algorithm proposed by Werder \textit{et
  al.}~\cite{werder:05} to impose non-periodic boundary conditions
(NPBC) on the MD system. Details of the algorithm can be found
in~\cite{werder:05}. 

\section{Results}
\label{sec:results}

We first consider the flow past a CNT embedded normal to the flow
direction in order to compare our results with~\cite{werder:05}. We
then discuss the flow through a short CNT embedded parallel to the
flow direction.

\subsection{Flow past a nanotube}

We apply the hybrid LB-MD algorithm to the case of the flow of argon
around a CNT centered along the z-axis within a $30 \times 30 \times
4.254$ nm$^3$ domain $\Omega$. The CNT is of chirality $(16,0)$ with a
radius of $r=0.625$ nm. We choose the density of argon
$\rho_{Ar}=1008$ kgm$^{-3}$ and the temperature $T=215$ K. This
corresponds to the dimensionless state point $(T^*,\rho^*)=(1.8,0.6)$
where $T^*=k_B T \epsilon^{-1}_{ArAr}\mathcal{A}$ and
$\rho^*=\rho\sigma^3_{ArAr}m_{Ar}^{-1}\mathcal{A}$. We let $k_B$ be
the Boltzmann constant, $m_{Ar}=0.03994$ kgmol$^{-1}$ the atomic mass
of argon, and $\mathcal{A}$ Avogadro's number.

The MD subdomain size is $10 \times 10 \times 4.254$ nm$^3$ centered
around the CNT which is subdivided into $20 \times 20 \times 1$
sampling cells where $6465$ argon atoms are initially equilibrated for
$0.2$ ns. The width of the strip around the boundary on which both MD
and LB are simulated is $\delta_s=2.5$ nm.

We consider an LB domain of size $60 \times 60 \times 1$ covering the
entire domain where lattice nodes are centered on the corresponding MD
sampling cells. 
The viscosity of the LJ fluid is a parameter of the LB model and set
to $\nu=0.745 \cdot 10^{-7}$ m$^2$s$^{-1}$~\cite{meier:04}. We have
performed a sensitivity analysis by increasing and decreasing the
viscosity by $5\%$ and found it to be a robust parameter with respect
to accuracy.
Dirichlet boundary conditions $u_\infty=u_x=100$ ms$^{-1}$ are imposed
at the inlet $x=0$ nm and outlet $x=30$ nm. This high velocity is
chosen to reduce the number of sampling iterations. The temperature is
controlled by using a Berendsen thermostat~\cite{berendsen:84} with a
time constant $\tau_T=0.1$ ps. We apply it cell-wise in all directions
in the boundary cells where the velocity is prescribed and only in the
z-direction in other cells.
The hybrid model is run for $100$ cycles which consists of running the
LB simulation for $7$ ns ($15000$ iterations) followed by an MD step
equilibrating for $0.2$ ns ($20000$ iterations) and sampling for
$t_{s}=0.4$ ns. We discuss below how $t_s$ affects the convergence of
the results.

\begin{figure}
\begin{center}
\begin{tabular}{m{7cm}m{7cm}}
\epsfig{file=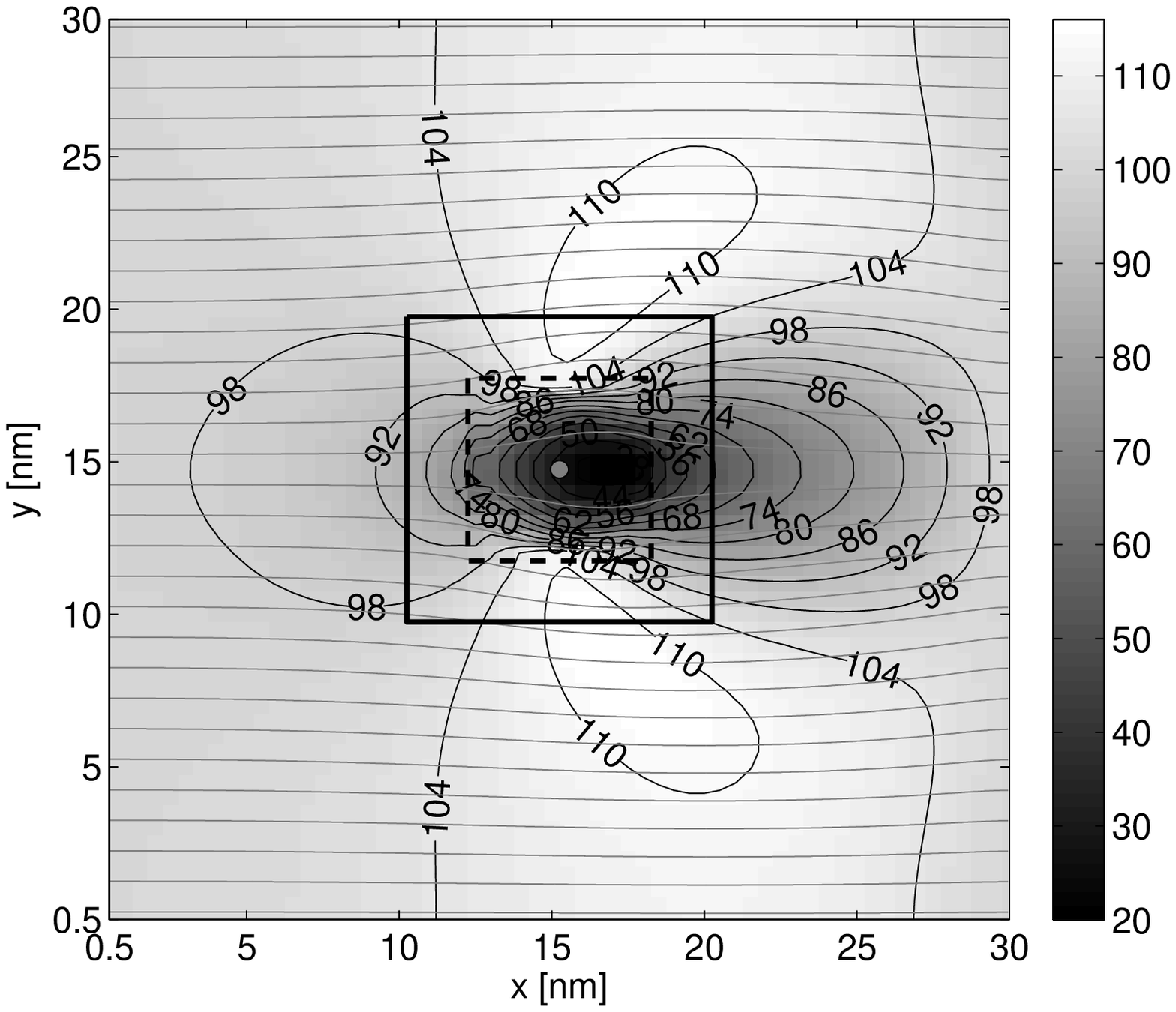,width=7cm} & \multirow{2}{*}{\epsfig{file=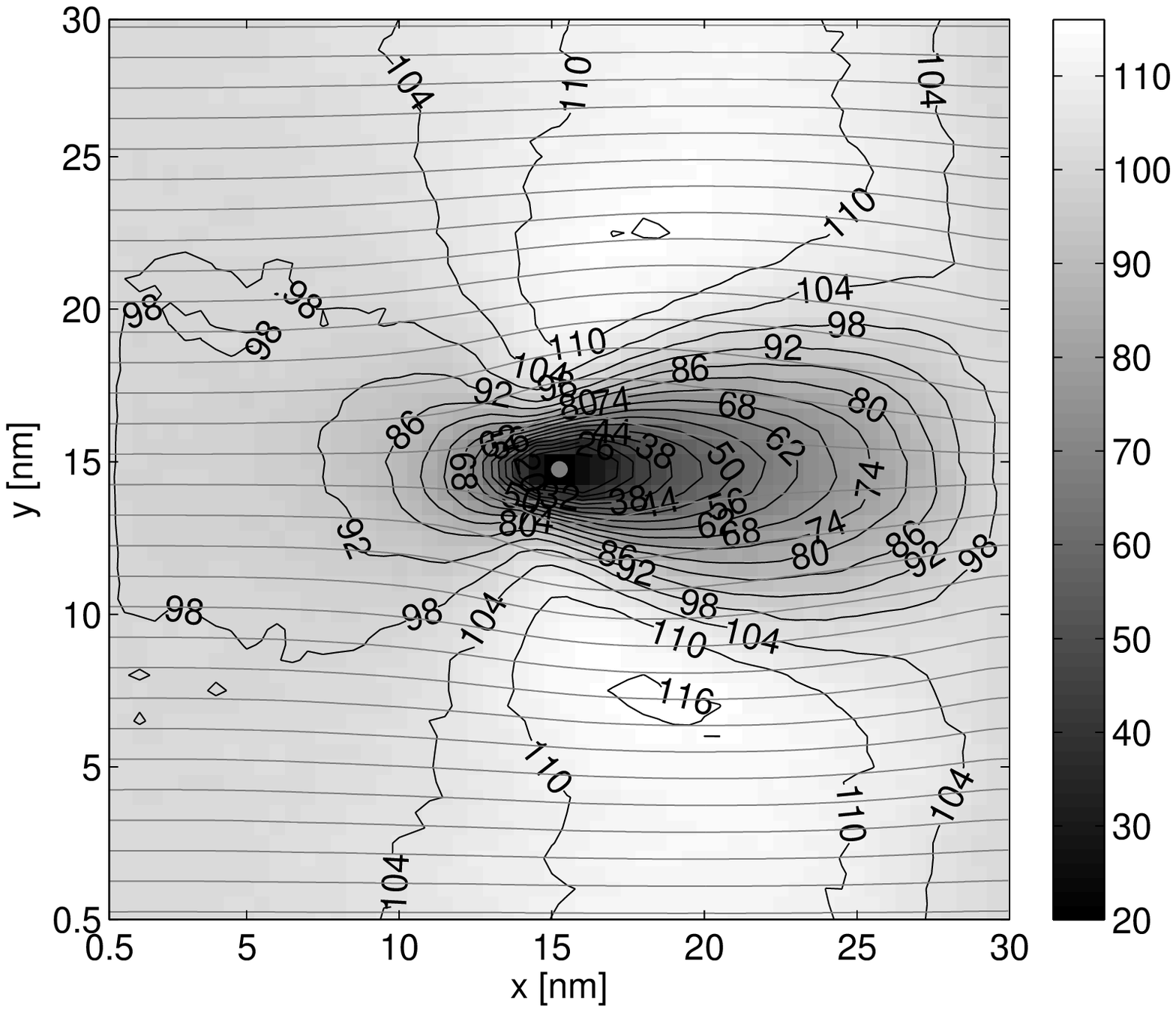,width=7cm}} \\
\epsfig{file=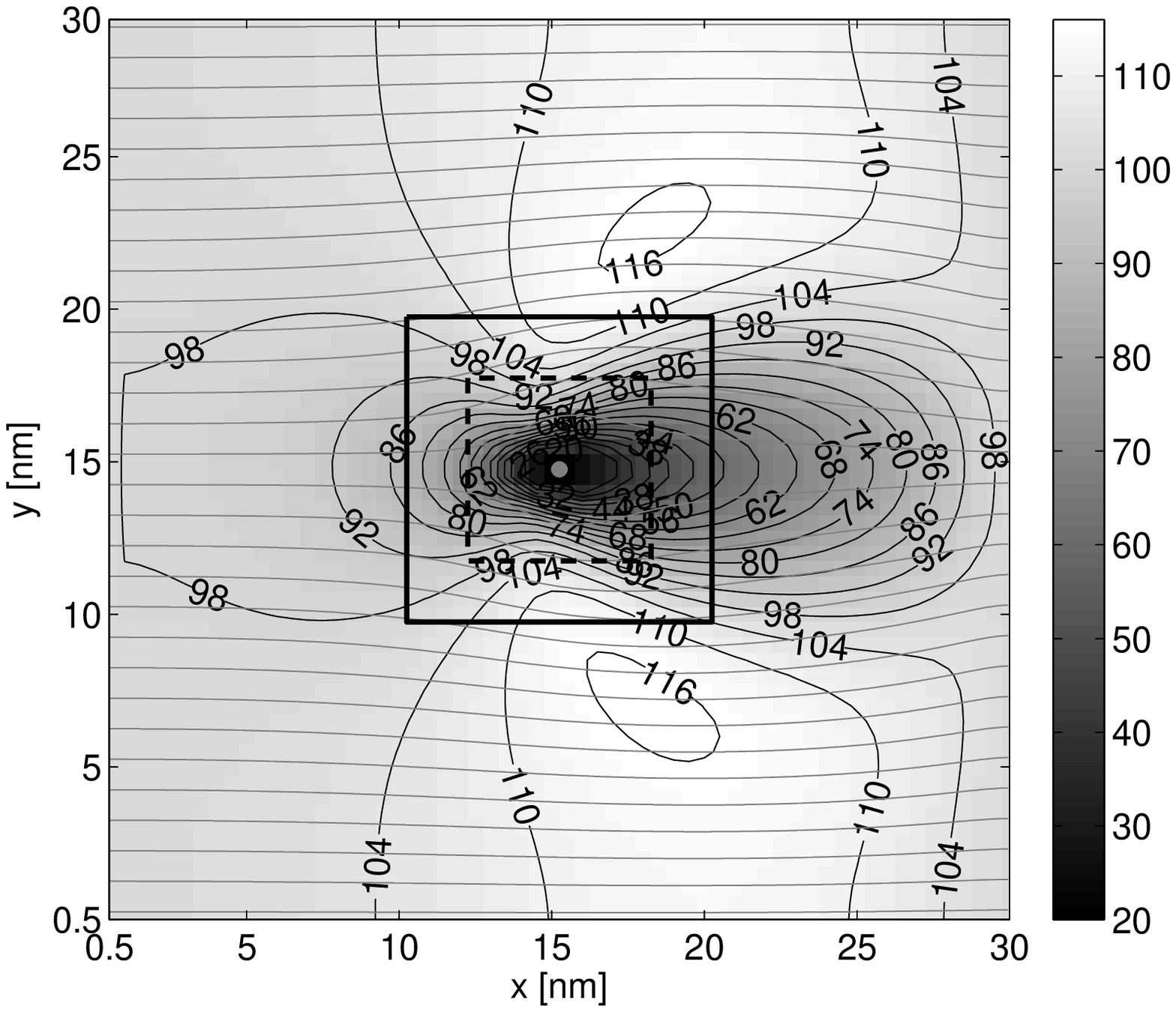,width=7cm} & \\
\end{tabular}
\end{center}
\caption{(left) Converged hybrid and (right) reference MD solutions of
  a flow past a CNT. (top) Velocities and (bottom) velocities and
  velocity gradients are enforced. The colorscale and contour lines
  depict the norm of the velocity expressed in ms$^{-1}$. Gray lines
  are streamlines. The thick squares represent (solid) the boundary of
  the MD subdomain and (dashed) the boundary of the overlapping
  region.}
\label{fig:nano}
\end{figure}

\Fig{fig:nano} shows a comparison between hybrid converged solutions
and an MD reference solution over the entire domain. The latter
involves $58198$ argon atoms and the temperature is controlled as in
the MD subdomain. The reference MD velocity field is sampled over $20$
ns. A qualitative match with the reference MD solution is obtained
when using the VC approach. The match between the hybrid solution and
the reference MD solution becomes quantitative when using the VGC
method. This is due to the fact that velocity gradients are not let
free but implicitly imposed. A consequence of this is that
discrepancies in the wake and on the sides of the CNT that appears
when using the VC approach, and also reported in~\cite{werder:05}, are
not observed when using the VGC method.

\begin{figure}
\begin{center}
\begin{tabular}{cc}
\epsfig{file=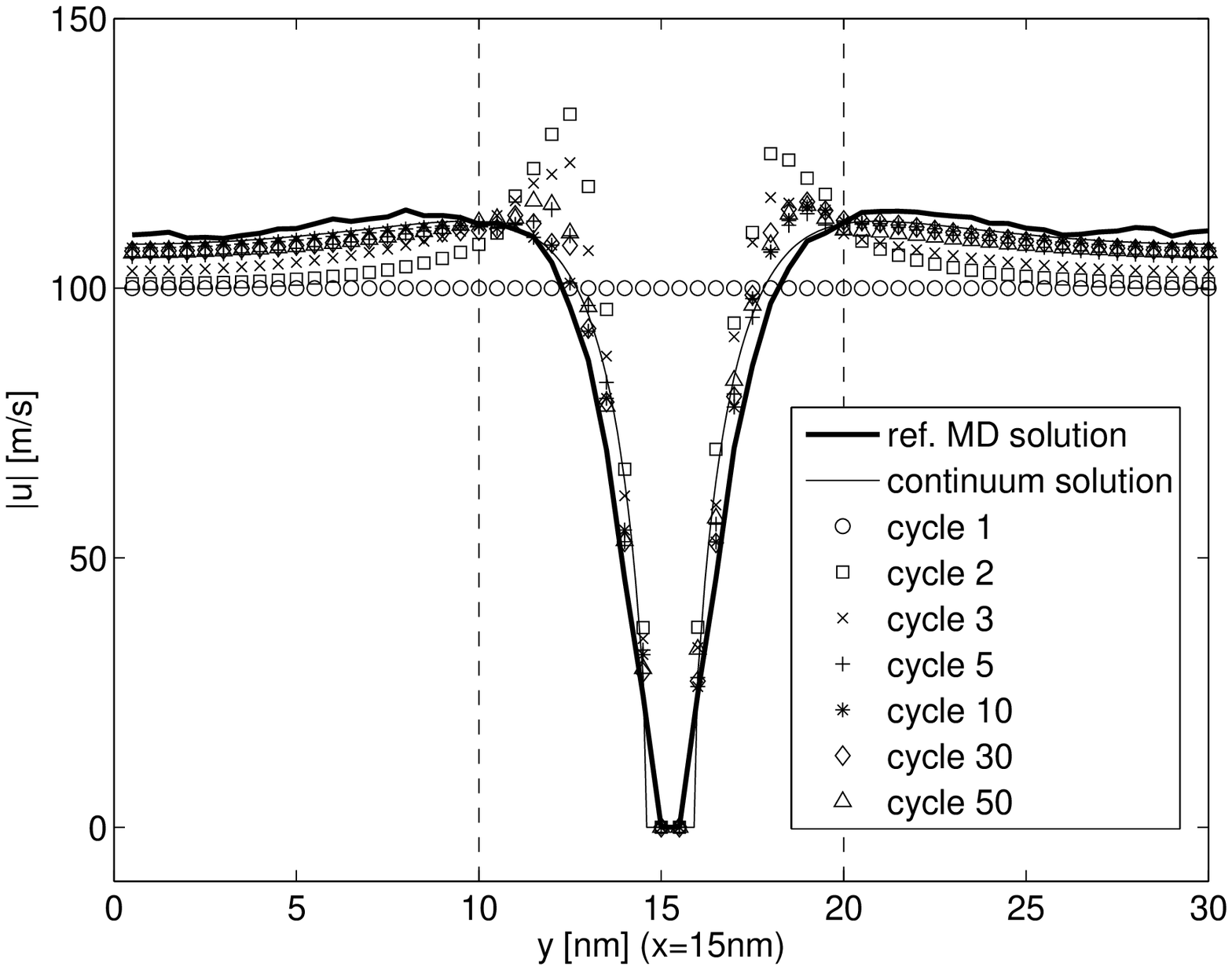,width=7cm} &
\epsfig{file=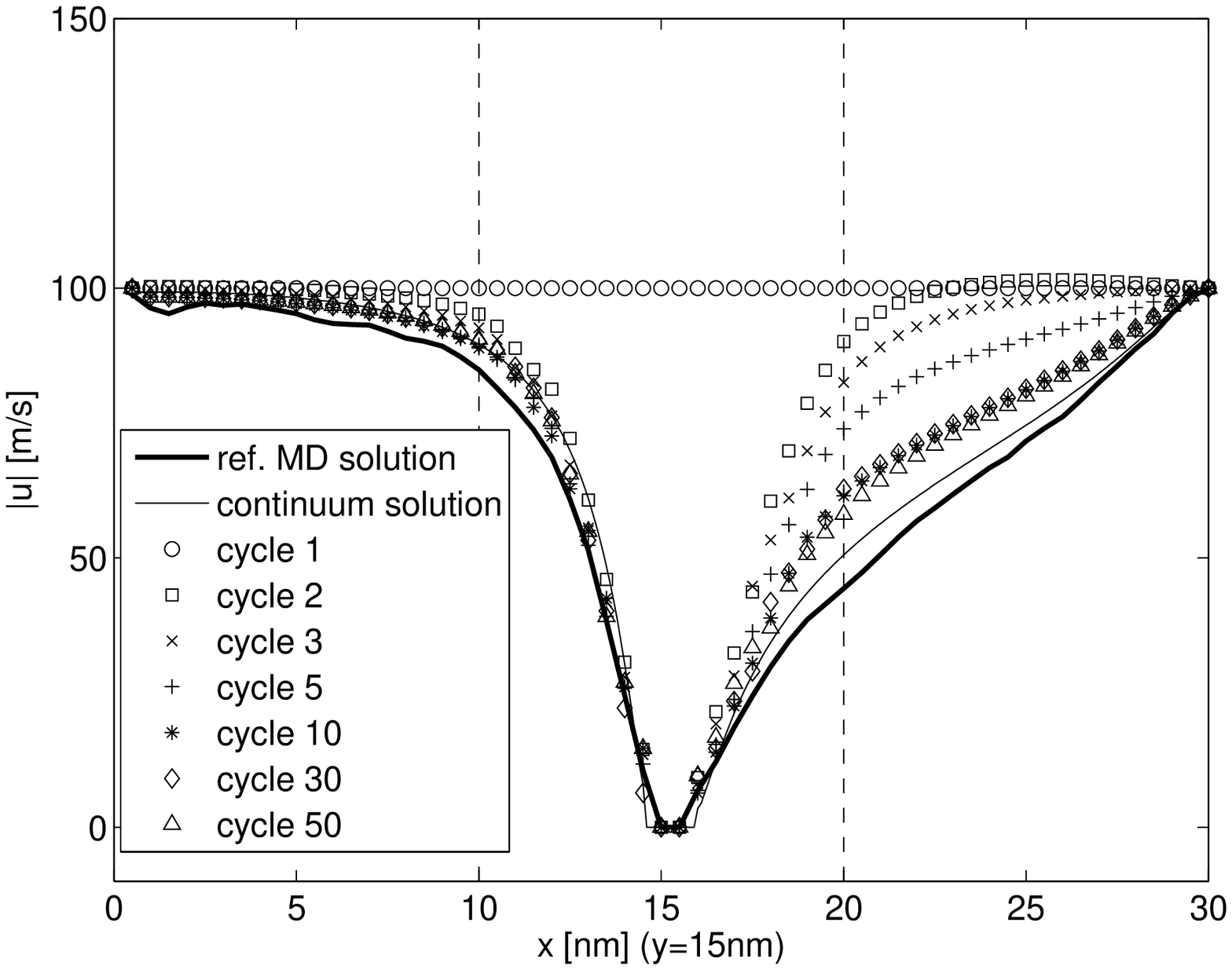,width=7cm} \\
\epsfig{file=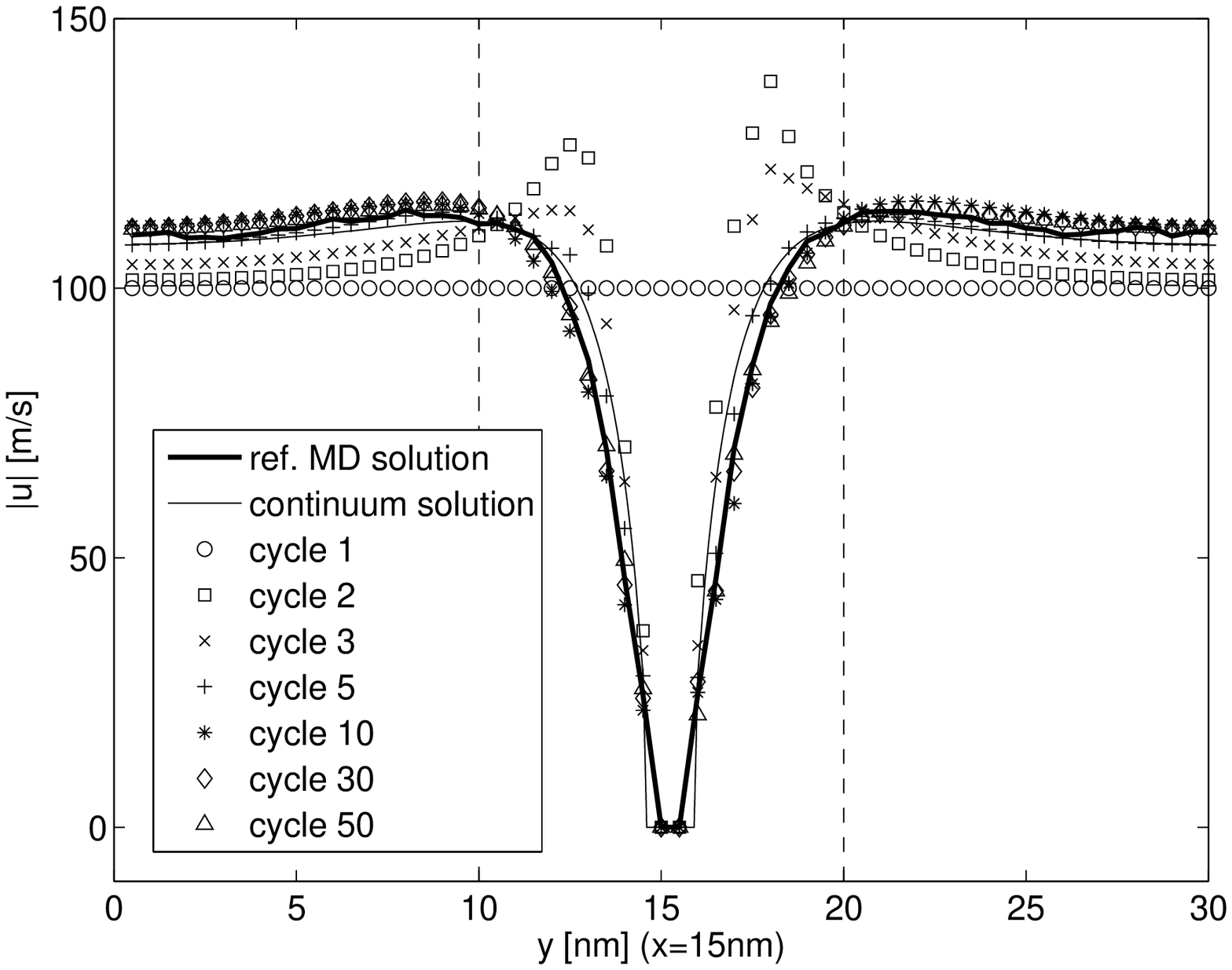,width=7cm} &
\epsfig{file=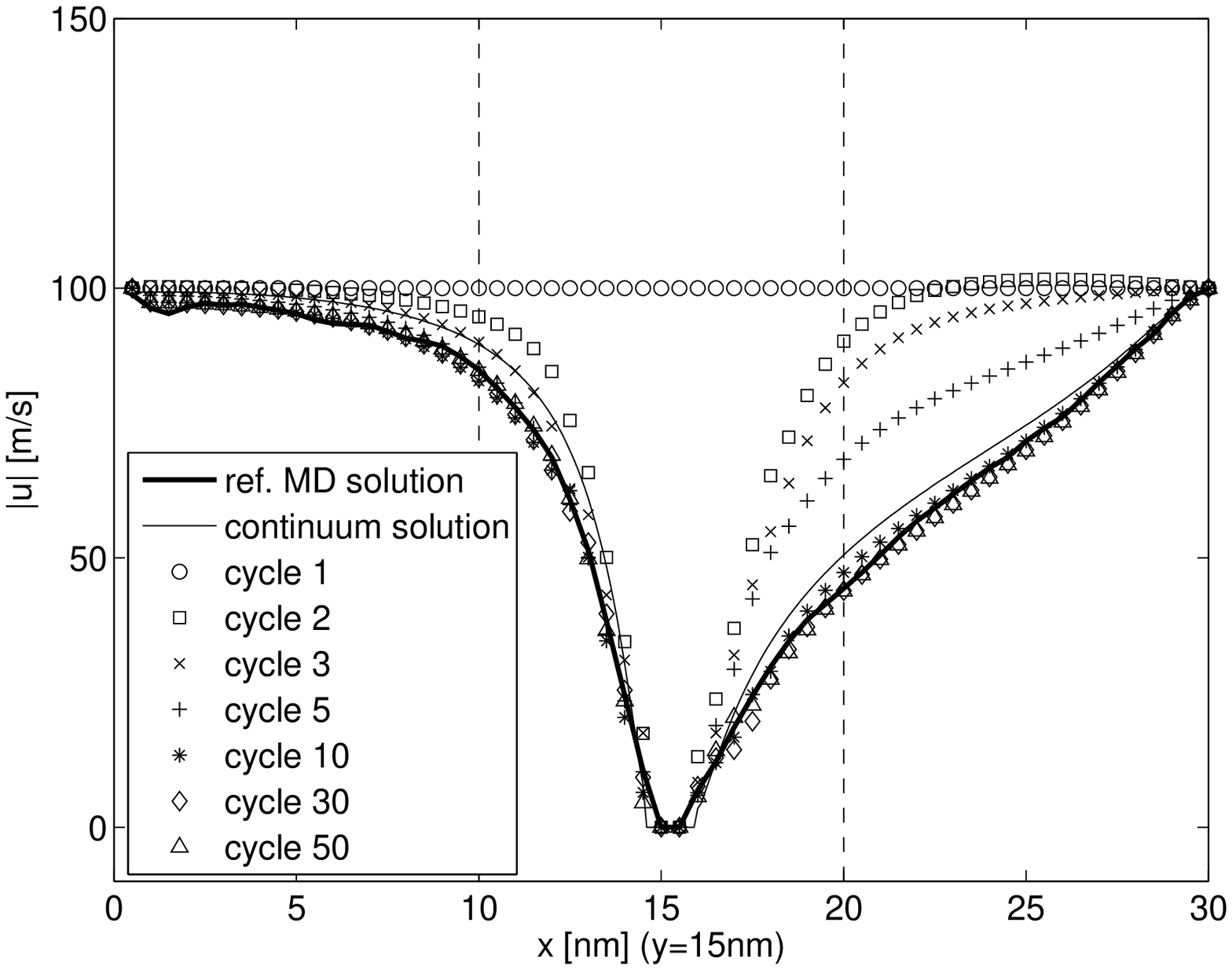,width=7cm}
\end{tabular}
\end{center}
\caption{Evolution of the hybrid profiles of the norm of the velocity
  in the planes located (left) at $x=15$ nm and (right) at $y=15$
  nm. (top) Velocities and (bottom) velocities and velocity gradients
  are enforced. The solid thick line is the reference MD solution
  whereas the solid line is the continuum solution where no-slip
  boundary condition is considered around the CNT. Hybrid solutions
  are plotted after various number of cycles. Dashed lines represent
  the boundaries of the MD subdomain.}
\label{fig:velProf}
\end{figure}

\Fig{fig:velProf} shows the evolution of the norm of the velocity
along the $y=15$ nm and $x=15$ nm plane. The system starts from a
constant initial condition $u_x=u_\infty=100$ ms$^{-1}$. This
unphysical state leads to the formation of high velocity regions
around the CNT. They then slowly spread out on the sides of the tube
and increase the side velocity. Considering the VGC approach, we
observe that the velocity upstream and downstream adjusts smoothly to
the reference solution. The situation is different when employing the
VG method. Indeed the hybrid velocity profiles show deviations from
the MD reference solution especially downstream the CNT. It is worth
noting that we get identical results when initializing the system with
initial conditions closer to the steady solution.

In agreement with~\cite{walther:04}, we observe that argon flow past  a CNT
normal to the flow results in a vanishing  velocity at the surface of
the CNT. In this particular configuration, the velocity boundary condition around the tube can therefore
be approximated by a no-slip condition allowing us to solve the NS
equations. \Fig{fig:velProf} shows LB velocity profiles where the CNT
is modeled off-lattice using an Immersed Boundary
technique~\cite{dupuis:06b}. We will see however that this 
approximation is not applicable when we consider a  CNT
with a different orientation.

We quantify the convergence towards the reference MD solution by
defining an error $e^j$ between the hybrid solution at cycle $j$ and
the reference MD solution as
\begin{equation}
e^j=\frac{1}{N} \sum_{k \in \Omega}
\frac{|\mathbf{u}_k^j-\mathbf{u}_{k,MD}|}{u_\infty} 
\label{eq:err}
\end{equation}
where $N$ is the number of cells in $\Omega$, $\u_k^j$ and
$\mathbf{u}_{k,MD}$ are respectively the hybrid and reference MD
velocities at cycle $j$ in the cell $k$.

\begin{figure}
\begin{center}
\begin{tabular}{cc}
\multicolumn{2}{c}{\epsfig{file=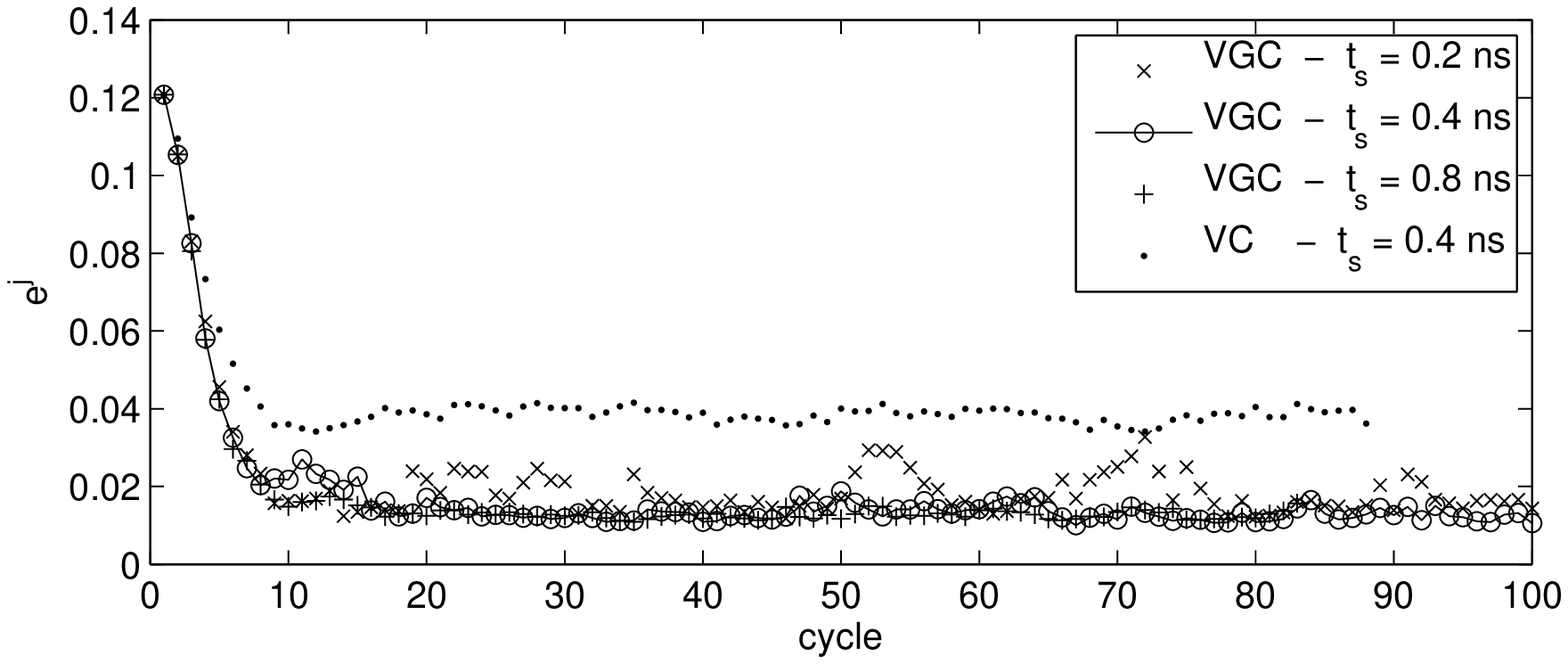,width=11cm}} \\
\epsfig{file=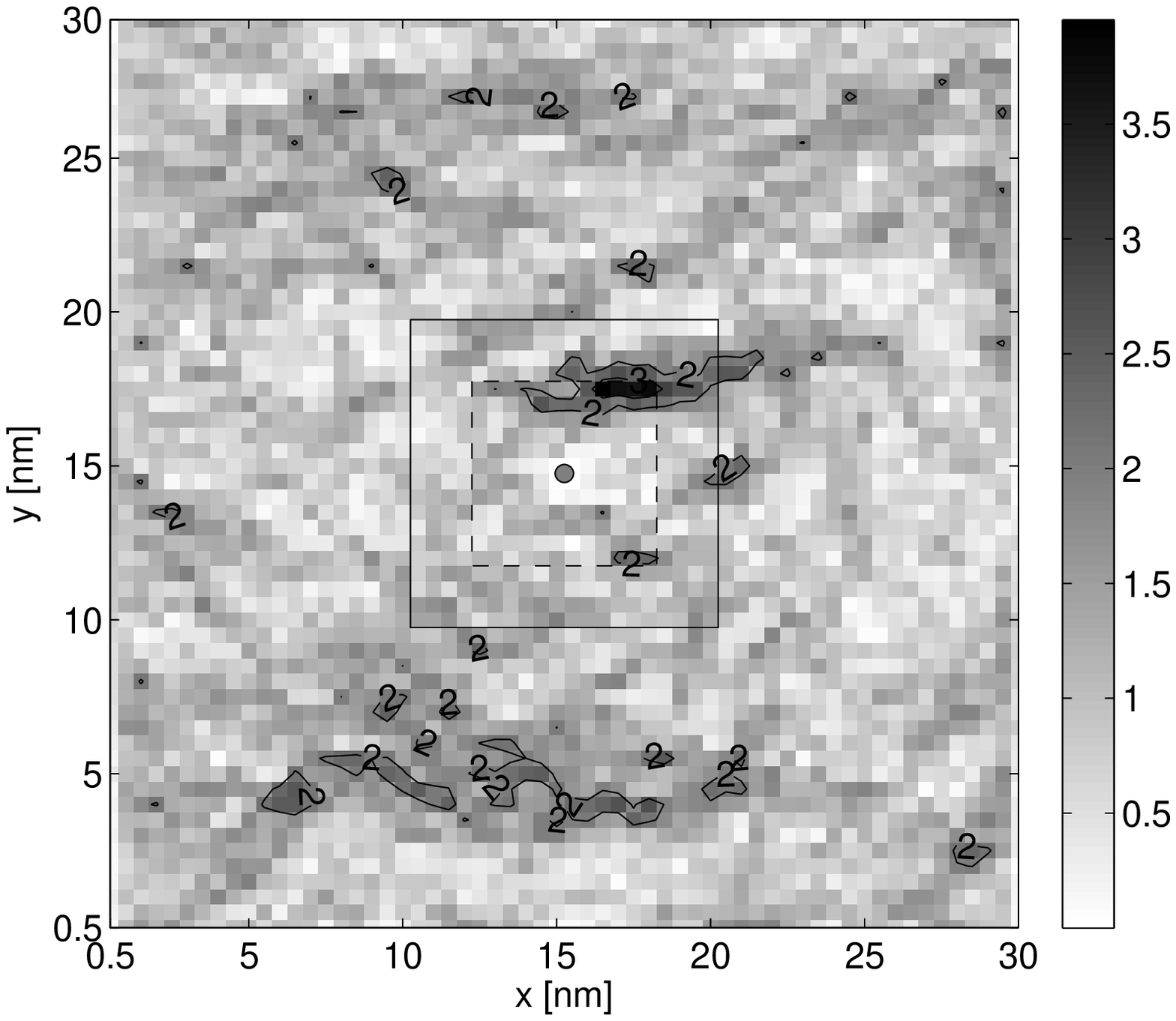,width=7cm} & \epsfig{file=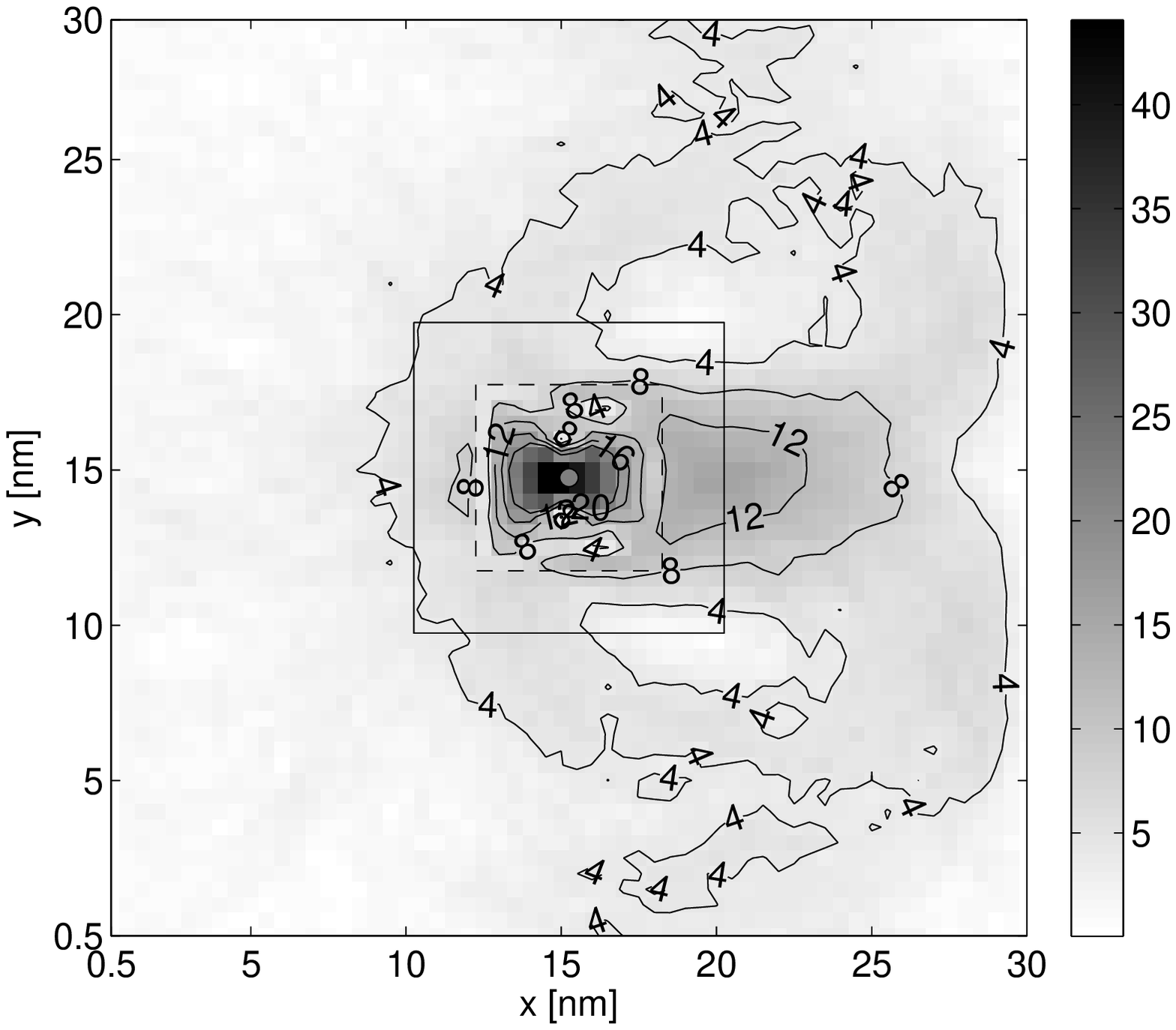,width=7cm} 
\end{tabular}
\end{center}
\caption{(top) Evolution of the error $e^j$ between the hybrid
  solution and the reference MD solution. Different sampling time
  $t_s$ and coupling techniques are considered. (bottom) Point-wise
  error according to \eq{eq:err} shown in percentage and averaged
  between cycle $50$ and $100$. The sampling time $t_s=0.4$ ns. (left)
  The VGC and (right) the VC techniques are used.}
\label{fig:error}
\end{figure}

\Fig{fig:error} shows the time evolution of the error which rapidly
decays during the first $10$ cycles. The error then fluctuates around
an average value which is a function of $t_s$. Considering the VGC
approach, we measure an average error between cycle $50$ and $100$ of
$1.3 \%$ for $t_s=0.4$ ns and $t_s=0.8$ ns and an average error of
$1.9 \%$ for $t_s=0.2 $ ns. We observe in \fig{fig:error} that
considering short $t_s$ leads to undesirable fluctuations whereas long
$t_s$ decreases the computational efficiency of the model. An optimum
is found to be $t_s=0.4$ ns assumed hereafter if not otherwise
specified. \Fig{fig:error} also shows the time evolution error when
using the VC method. The average error between cycle $50$ and $100$ is
of $3.8 \%$ and is comparable to the the error evolution reported
in~\cite{werder:05}.

The point-wise error between the hybrid and the reference MD solutions
is depicted in \fig{fig:error}. Using the VGC technique leads to
a localized and maximum $3.9 \%$ error region showing only on one side
of the CNT. Everywhere else the error is up to $2.5 \%$. The error is
higher when using the VC approach where we observe a large error of
$40 \%$ close to the CNT and up to $12\%$ in the wake.

The supercritical state point $(T^*,\rho^*)=(1.8,0.6)$ is associated
with less structural correlations than in a liquid state. In order to
assess their effect on the convergence we have considered the liquid
state point $(131 K, 1361$ kgm$^{-3})=(1.09,0.81)$ related to the
kinematic viscosity $\nu=1.42 \cdot 10^{-7}$
m$^2$s$^{-1}$~\cite{meier:04}. For similar parameters as for the state
points $(1.8,0.6)$, we measure an error of $2.0 \%$ between the hybrid
solution and the associated reference MD solution.  The error is
slightly higher than the one obtained in the supercritical state and
indicates that the LB-MD model can also be applied when considering
liquid states.

\subsection{Flow through a nanotube}

We now consider the flow through a CNT driven driven by a constant
velocity $u_x=100$ ms$^{-1}$ enforced at $x=0$ nm and $x=28$. The CNT
is centered along the x-axis in a domain of size $28 \times 16 \times
16$ nm$^3$ and it is of chirality $(16,0)$ with a length $l=2.34$
nm. We choose the dimensionless state point $(T^*,\rho^*)=(1.8,0.6)$. 

The MD subdomain size is $14 \times 10 \times 10$ nm$^3$ centered
around the CNT and is subdivided into $14 \times 10 \times 10$
sampling cells. $21278$ argon atoms are initially equilibrated for
$0.2$ ns and $\delta_s=3$ nm. The temperature is regulated by applying
a Berendsen thermostat cell-wise in every direction within the
boundary cells.
We consider a $28 \times 16 \times 16$ LB domain covering the entire
domain where lattice nodes are centered in corresponding sampling
cells. MD and LB domains are coupled via the VGC approach. The hybrid
model is ran for $50$ cycles consisting of running the LB simulation
for $13.4$ ns ($15000$ iterations) followed by an MD step equilibrated
for $0.2$ ns and sampled for $t_{s}=0.4$ ns.

\begin{figure}
\begin{center}
\begin{tabular}{ccc}
\epsfig{file=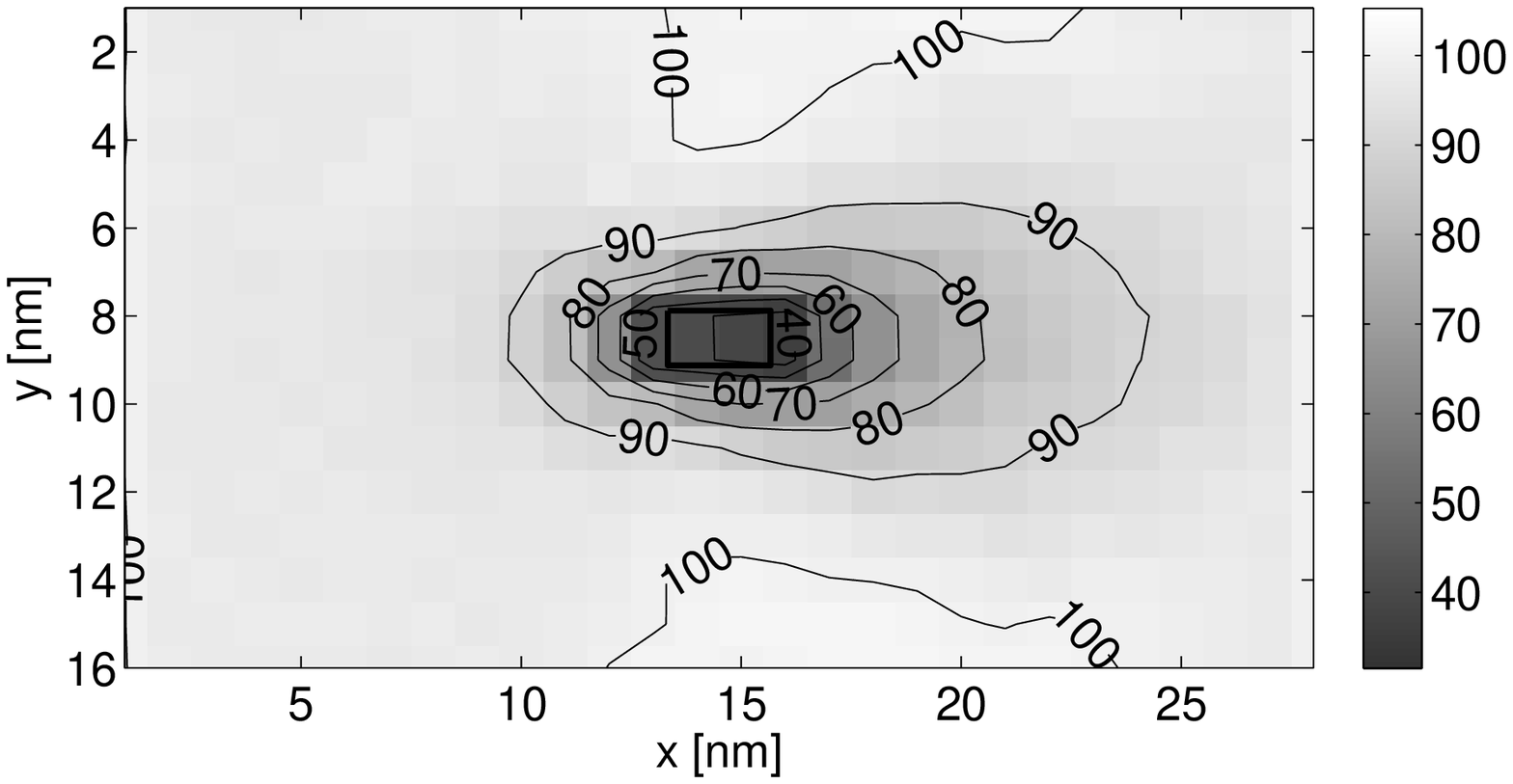,width=5cm} & 
\epsfig{file=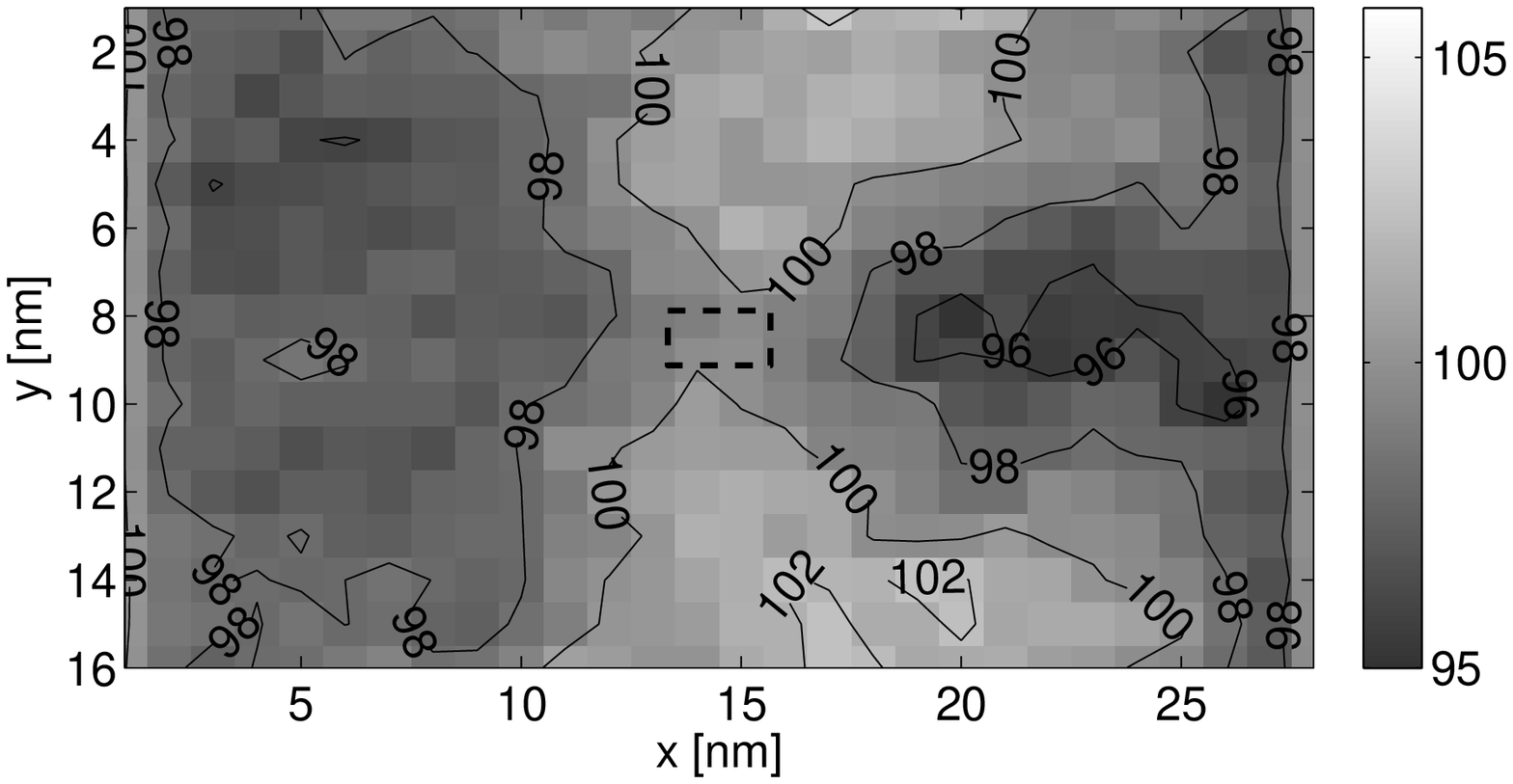,width=5cm} & 
\epsfig{file=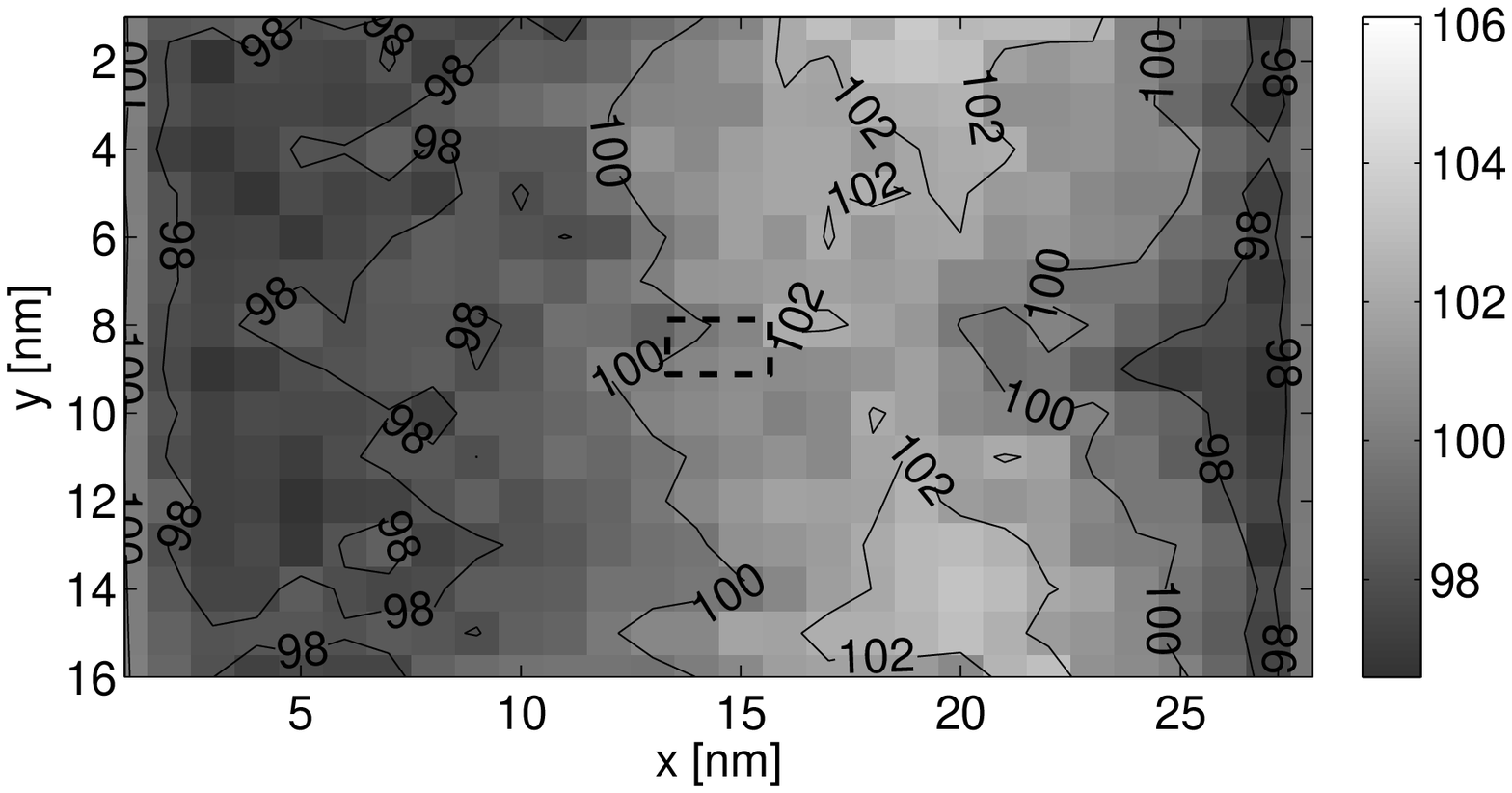,width=5cm} \\
\epsfig{file=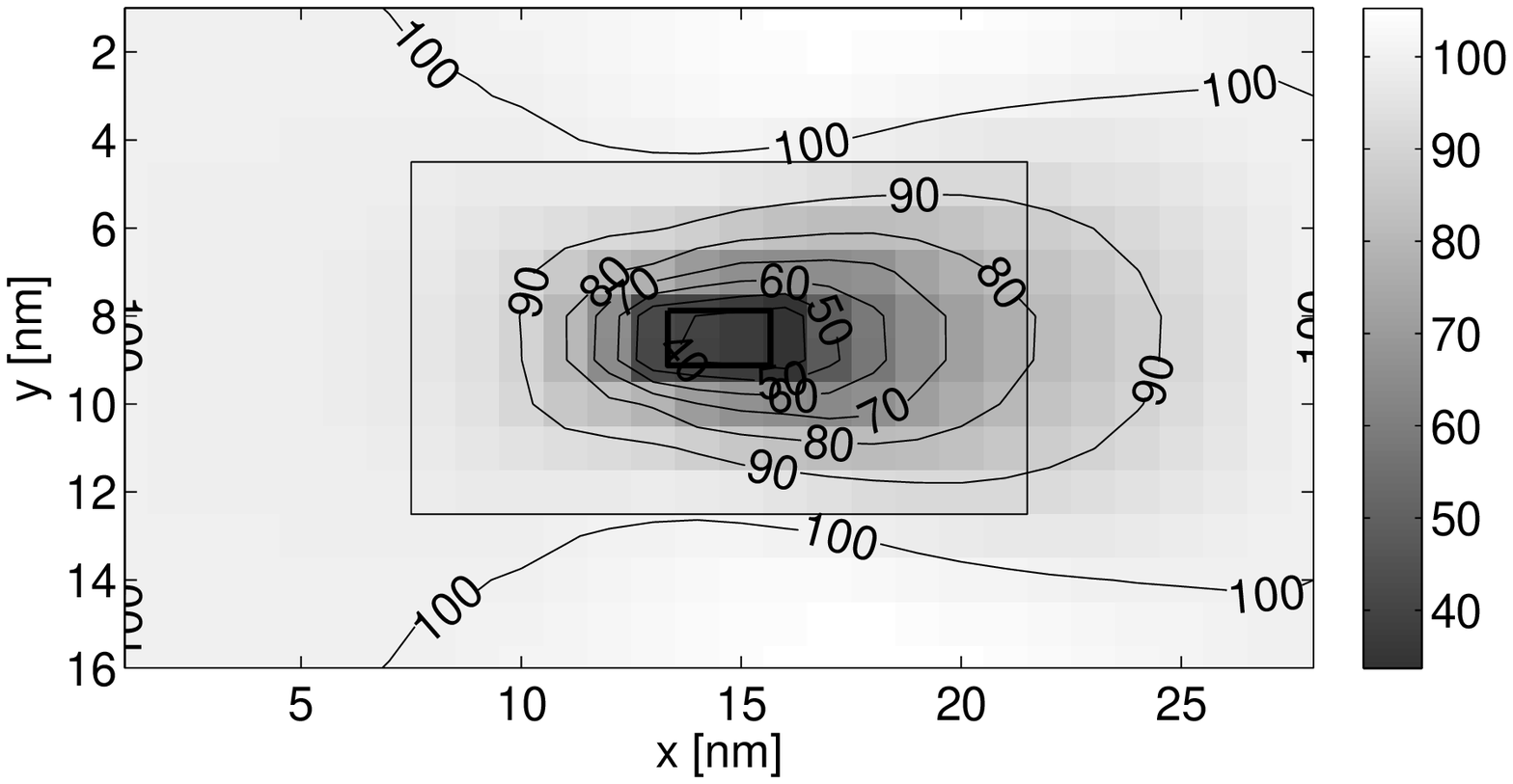,width=5cm} & 
\epsfig{file=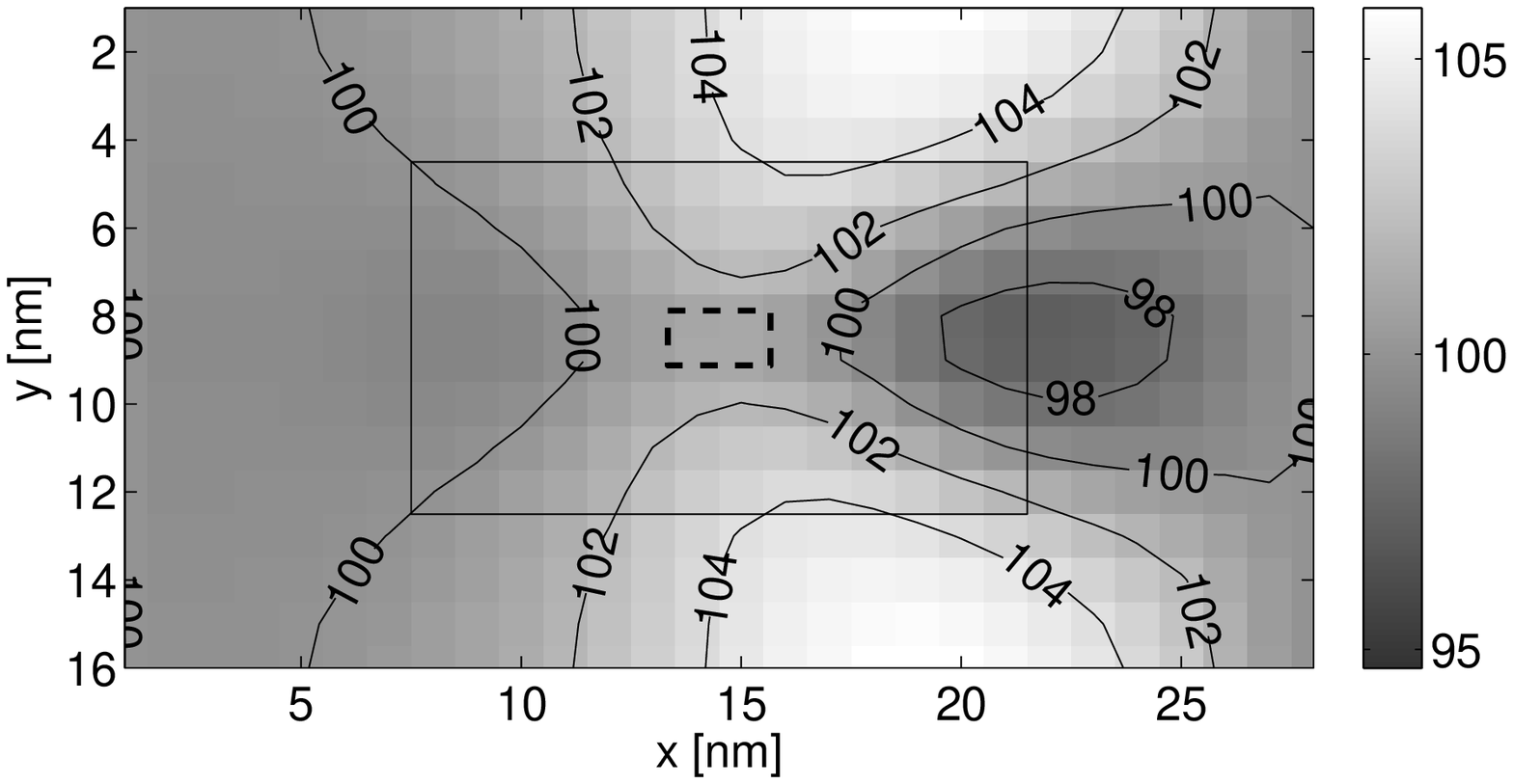,width=5cm} & 
\epsfig{file=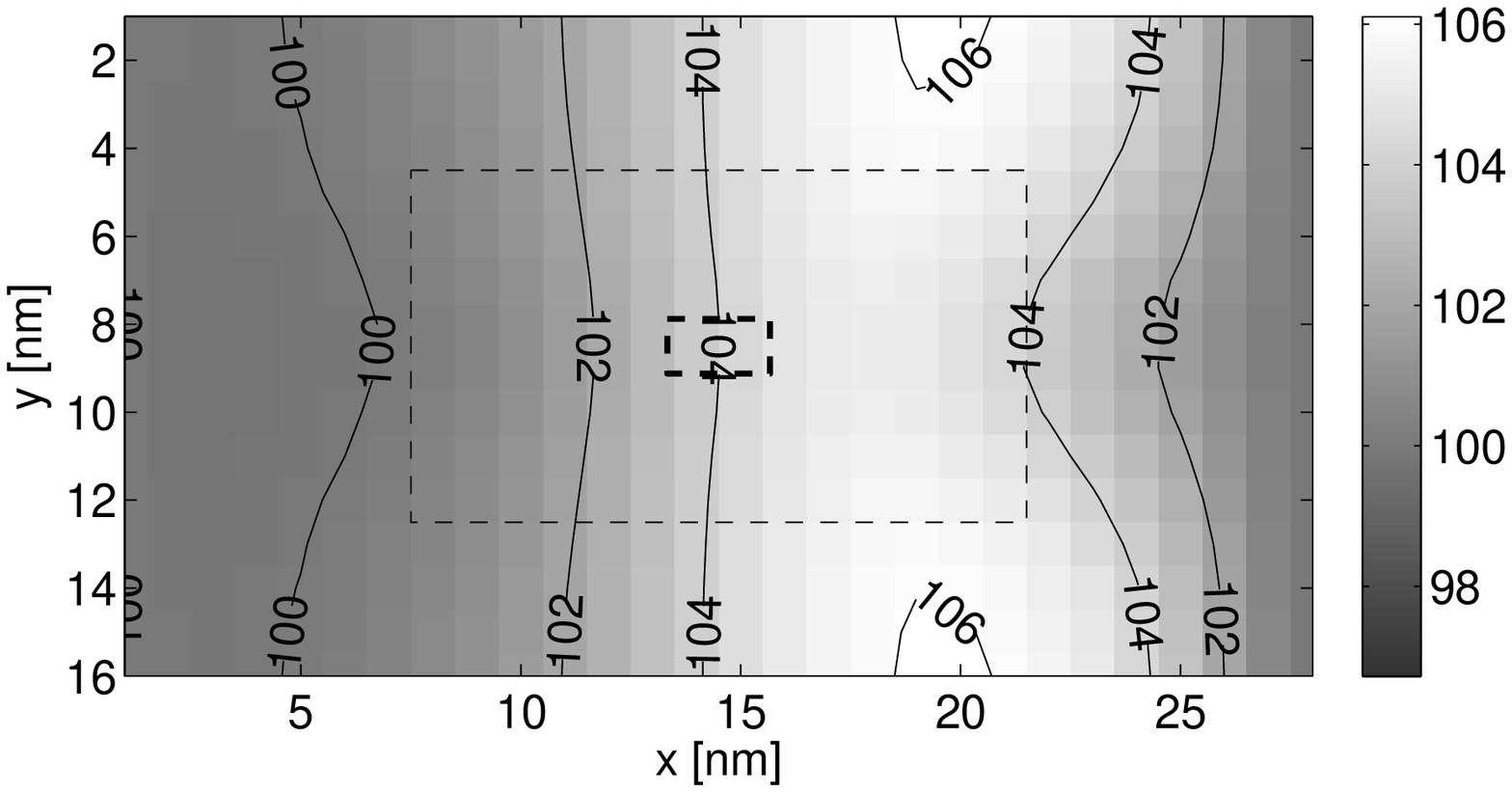,width=5cm} \\
$z=8$ nm & $z=4$ nm  & $0$ nm \\
\end{tabular}
\end{center}
\caption{Reference MD (top) and hybrid (bottom) solutions of the flow
  through a CNT driven by a constant velocity $u_x=100$ ms$^{-1}$
  enforced at $x=0$ nm and $x=28$ nm. The norm of the velocity is
  plotted. Each column corresponds to a different $x-y$ plane at $z=8,
  4$ and $0$ nm. Contour lines are plotted and expressed in
  ms$^{-1}$. Bold and thin rectangles depict the CNT and the MD
  subdomain respectively.}
\label{fig:nano3D}
\end{figure}

\Fig{fig:nano3D} shows a comparison between the hybrid and the
reference MD solution. The latter consists of $108943$ argon atoms
sampled for $20$ ns where thermal boundary conditions are as in the MD
subdomain. There is an overall good quantitative agreement between the
solutions. We observe regions above and under the CNT where the hybrid
velocity is slightly faster than in the reference MD solution. The
difference is at most $4\%$. We quantify the overall agreement by
measuring average errors following \eq{eq:err} over different regions
and during the last $30$ cycles and obtain a global error $e_g=2.6
\%$ over the whole domain and a local error $e_l=2.1\%$ in the tube.

The rather small diameter of the CNT leads to a higher argon density
within the tube. We measure a $7\%$ increase that can not be described
by the incompressible NS equations. This issue could however be
alleviated by considering a compressible LB model~\cite{briant:04}.

In this conformation, the velocities around the CNT are non-zero. The
corresponding continuum velocity boundary condition is unknown and a
function of the system parameters~\cite{walther:04}. A continuum
solution approximating the reference solution is therefore unfeasible.

The computational efficiency of the LB-MD model is estimated by
comparing the time needed to compute one iteration of the reference
and hybrid solution, respectively. We measure $t_{ref}=0.56$ s and
$t_{hyb}=0.08$ s in the case of the flow past a CNT. The hybrid
solution is computed $R=0.56/0.08=7$ times faster than the reference
solution. We get a ratio $R=1.04/0.43=2.4$ in the case of a flow
through a CNT. Ratios $R$ are of the same order as the volume ratio
between the MD domain and subdomain. Small systems have been chosen
here in order to compare hybrid to reference solutions. Considering
larger systems would exhibit much higher ratios. For example,
nanodevices ($\propto 10^3$ nm$^3$) embedded in microscale systems ($
\propto 1$ $\mu$m$^3$) would lead to ratios of the order of $10^6$.

\section{Conclusion}
\label{sec:conclusion}

We have presented a hybrid model coupling a Lattice Boltzmann solution of the 
incompressible Navier-Stokes equations  to a Molecular Dynamics 
Simulation of a dense fluid, using a Schwarz domain decomposition technique.
The two descriptions are coupled via an exchange of velocities (VC) or velocity 
gradients (VGC).
The applicability of the method was demonstrated in flows of liquid argon
past Carbon Nanotubes normal and aligned with the flow velocity.

We have first computed the flow past a CNT with an axis normal 
to the flow velocity. Using the VGC approach we observed quantitative agreement
between the hybrid and the reference MD solutions with an average
error of $1.3\%$ that provides an improvement over 
previously reported results \cite{werder:05} for the same configuration.
We attribute  this improvement to the fact that we enforce
velocity within a region rather than along a strip implying  that 
velocity gradients and in turn shear forces 
are implicitly exchanged between the two descriptions. We also show that the
velocity vanishes around the tube in this conformation making a
continuum solution a good approximation of the reference
solution. This is not the case however  when considering the flow
through a CNT. The transport of argon in the CNT is captured within
$2.1\%$ with our LB-MD model. To the best of our knowledge, this is the first time
that fully 3D hybrid simulations have been reported in the literature.

In addition of being a tool to help investigating multi-scale physics
by considerably reducing an otherwise prohibitive computational time,
this approach largely extends the applicability of LB models by
providing us with fully microscopic boundary conditions.

\section*{Acknowledgment}

We wish to thank P.G.~Gonnet for many helpful technical discussions.

\end{document}